\documentclass[draft]{agujournal}

\usepackage{amsmath,amssymb,latexsym}  
\usepackage{bm}
\usepackage{mathtools}
\usepackage{textcomp}
\usepackage[english]{babel}
\usepackage{hyperref}

\journalname{JGR-Planets}

\begin{document}

%
%


\title{A past lunar dynamo thermally driven by the precession of its inner core}


%
%




 \authors{
Christopher Stys\affil{1} \hspace*{0.075cm} and Mathieu Dumberry\affil{1}}
\affiliation{1}{Department of Physics, University of Alberta, Edmonton, Alberta, Canada.}






\correspondingauthor{Mathieu Dumberry}{dumberry@ualberta.ca}




\begin{keypoints}
\item Viscous heating from the differential precession between the Moon's fluid and solid cores was high enough in the past to power a dynamo.
\item The surface magnetic field that it produces is of the order of a few microTeslas, compatible with the lunar field recorded after 3 Ga. 
\item The associated heat flux at the inner core boundary is an important contribution to the thermal evolution of the lunar core.
\end{keypoints}

%
%


\begin{abstract}
The Cassini state equilibrium associated with the precession of the Moon predicts that the mantle, fluid core and solid inner core precess at different angles.  We present estimates of the dissipation from viscous friction associated with the differential precession at the core-mantle boundary (CMB), $Q_{cmb}$, and at the inner core boundary (ICB), $Q_{icb}$, as a function of the evolving lunar orbit.  We focus on the latter and show that, provided the inner core was larger than 100 km, $Q_{icb}$ may have been as high as $10^{10}-10^{11}$ W for most of the lunar history for a broad range of core density models. This is larger than the power required to maintain the fluid core in an adiabatic state, therefore the heat released by the differential precession at the ICB can drive a past lunar dynamo by thermal convection.  This dynamo can outlive the dynamo from precession at the CMB and may have shutoff only relatively recently.  Estimates of the magnetic field strength at the lunar surface are of the order of a few $\mu$T, compatible with the lunar paleomagnetic intensities recorded after 3 Ga.  We further show that it is possible that a transition of the Cassini state associated with the inner core may have occurred as a result of the evolution of the lunar orbit. The heat flux associated with $Q_{icb}$ can be of the order of a few mW m$^{-2}$,  which should slow down inner core growth and be included in thermal evolution models of the lunar core.
\end{abstract}

{\bf Plain language summary:} While the Moon today no longer has a large scale magnetic field generated by dynamo action in its small iron core, magnetic studies on lunar rocks collected during the Apollo missions suggest that it did in the past.  However, the mechanism responsible for this dynamo is still debated.  In this study, we investigate whether the precession motion of the Moon may have been capable to sustain a past lunar dynamo.  The mantle, fluid core, and solid inner core of the Moon precess at different angles today.  These precession angles were larger in the past when the Moon was orbiting closer to Earth. We calculate the dissipation generated by the viscous friction from the differential precession at the boundaries between the fluid core and the mantle (CMB) and between the fluid and solid cores (ICB).  We focus of the latter and show that dissipation at the ICB in the past was high enough to drive thermal convection and sustain a dynamo.  A lunar dynamo driven by this mechanism is long-lived and may have shutoff as late as 1 billion years ago and perhaps even more recently.  

\section{Introduction}
\label{sec:intro}

The Moon does not currently possess a global magnetic field generated by dynamo action.  However,  remanent magnetization measured in the crust by satellites \cite[e.g.][]{mitchell08,purucker10} and on lunar rock samples collected during the Apollo missions \cite[e.g.][]{weiss14} both suggest that a dynamo was operating in the past.  Paleomagnetic analyses on Apollo samples indicate that a dynamo characterized by high surface intensities of several tens of $\mu$T to perhaps as high as 120 $\mu$T operated early in the lunar history between about 4.25 and 3.56 Ga \cite[][]{garrickbethell09, garrickbethell17,cournede12,shea12,tikoo12,suavet13}, although the accuracy of the very high paleointensity values have been called into question \cite[e.g.][]{lepaulard19}. This high-field epoch is followed by weaker paleointensities below 4 $\mu$T from 3.2 Ga onwards \cite[][]{tikoo14,tikoo17}.  The lunar paleomagnetic record is more spotty in this weak-field epoch and we do not know with high accuracy when the lunar dynamo turned off, though there is good evidence that it may have persisted until as recently as 1 Ga \cite[][]{tikoo17,mighani20}.

No single dynamo mechanism has so far been shown to be capable of explaining both the large paleointensities of the high-field epoch and the longevity of the weak-field epoch.  Explaining the large surface field recorded during the high-field epoch is particularly challenging.  A long lasting dynamo driven by thermo-chemical convection in the liquid core may explain the paleomagnetic record of the low-field epoch \cite[][]{laneuville14,scheinberg15}, but not the large intensities of the high-field epoch \cite[][]{evans18}.  Other possibilities that have been suggested to explain parts of the paleomagnetic record include a short-lived early core dynamo following a mantle overturn event \cite[][]{stegman03}, a dynamo generated in a magma ocean at the base of the mantle \cite[][]{scheinberg18} and a mechanically forced dynamo induced either by impacts \cite[][]{lebars11} or mantle precession \cite[][]{williams01,dwyer11,cebron19,cuk19}.  

A lunar dynamo powered by precession motion is the scenario that we further explore in our study.  The basis for this idea stems from the rotational dynamics of the Moon which is characterized by a Cassini state, and in which the orbit normal and spin-symmetry axis remain coplanar with, and are precessing about, the normal to the ecliptic \cite[][]{colombo66,peale69}.  The precession is retrograde, with a present-day period of 18.6 yr.  Lunar Laser Ranging (LLR) observations \cite[e.g.][]{dickey94,williams01} indicate a present-day tilt of the spin-symmetry axis of $1.543^\circ$ with respect to the ecliptic, though this only applies for the solid outer shell of the Moon comprised of its mantle and crust.  The spin axis of the fluid core should also lie in the plane that defines the Cassini state, however we do not expect it to be aligned with the symmetry axis of the mantle.  This is because the amplitude of the pressure torque exerted by the pole-to-equator core-mantle boundary (CMB) flattening on a misaligned fluid core is too small for the fluid core to be locked into synchronous precession with the mantle \citep{poincare10,goldreich67}.  An equivalent and complementary way to express this is to consider the free precession period of a misaligned rotation vector of the fluid core with the symmetry axis of the CMB sustained by this pressure torque.  This rotation mode is referred to as the free core nutation (FCN) and its period, though not directly observed, is expected to be a few hundred years \cite[e.g.][]{viswanathan19}, much longer than the 18.6 yr period of forced mantle precession.  This implies that the fluid core does not have time to adjust to the precession motion of the mantle and is not efficiently entrained with it. The rotation vector of the fluid core should therefore remain in close alignment with the normal to the ecliptic, though its precise angle is not observed directly and is unknown.

The flow motion of a fluid core precessing at a different angle than the mantle cannot be represented by a simple rigid body rotation \cite[e.g.][]{tilgner15}.  This is, first, because a secondary flow must exist to satisfy the no-penetration condition at the elliptically shaped CMB, and second, because finite viscosity requires the additional presence of a boundary layer flow underneath the CMB.  Based on the amplitude of the differential velocity across this boundary layer at present-day, this boundary layer flow is expected to be turbulent \cite[][]{yoder81,williams01}.  An estimate of the viscous dissipation associated with this turbulent flow is inferred by LLR and, although the mechanical stirring is not sufficiently large to power a dynamo today, it may have been in the past, when the Moon was closer to Earth \cite[][]{williams01}.  Indeed, the past Cassini state of the Moon featured a larger mantle tilt angle \cite[][]{ward75} and a fluid core spin axis remaining closely aligned with the ecliptic normal \citep{meyer11}.  Estimates of the viscous dissipation associated with this past, larger differential velocity at the CMB suggest that it may have been sufficiently large to power a dynamo by mechanical forcing \cite[][]{dwyer11,cebron19,cuk19}.

It is uncertain whether an inner core is present at the centre of the Moon today.  Its presence has been suggested by seismic observations \cite[][]{weber11}, but this is not universally accepted \cite[e.g.][]{garcia11}.   If an inner core is present, like the mantle and fluid core, it is also forced to precess at a period of 18.6 yr and its spin-symmetry axis should also lie in the plane that defines the Cassini state \cite[][]{williams07}.  Its angle of tilt is unknown, but it is determined by the period of the free inner core nutation (FICN), a free mode of rotation similar to the FCN but associated with the inner core \cite[][henceforth referred to as DW16 and SD18, respectively]{dumberry16,stys18}.  In analogy with the FCN, the FICN period depends of the amplitude of the pressure torque applied on the inner core when its elliptical surface is misaligned with the spin axis of the fluid core. But in addition, the FICN period also depends of the gravitational torque between the misaligned figures of the inner core and mantle, and for the Moon, it is the latter that dominates [DW16, SD18]. 

Just as the tilt angle of the spin axis of the fluid core depends on the FCN period relative to the forcing period of 18.6 yr,  the tilt angle of the spin-symmetry of the inner core is set by how the period of the FICN compares with this forcing period [DW16, SD18].  The FICN period is not known, but for reasonable models of the interior density structure of the Moon it is expected to be in the range of 10 to 40 yr.  Because the FICN period is close to the 18.6 yr forcing period, a large tilt of the inner core with respect to the mantle can result by resonant amplification [DW16, SD18].  Since the spin axis of the fluid core is expected to remain closely aligned with the ecliptic normal, this implies that there could be a large misalignment between the rotation vectors of the fluid and solid cores, and a differential velocity at the inner core boundary (ICB).  If an inner core has been present for a good portion of the lunar past, then mechanical stirring in the fluid core caused by differential velocity at the ICB may have been capable of generating a dynamo.  Furthermore, the heat generated by viscous friction at the ICB is available to drive a dynamo by thermal convection.

The main objective of our study is to explore the latter scenario.  The possibility of dynamos driven by precession is an active area of research \cite[e.g.][]{tilgner05,lin16,cebron19}.  The morphology of the resulting magnetic field depends on whether the turbulent flows are confined to a boundary layer or whether they trigger large scale instabilities destabilizing the whole of the fluid core.  Here, we do not present a dynamical model of a dynamo sustained by core flows entrained by the precession of an inner core.  Instead, we focus on whether a thermally driven, convective dynamo may be powered by the heat released at the ICB from the viscous friction associated with a differentially precessing inner core.  To do so, we approach the problem from an energy balance perspective  \cite[e.g.][]{nimmo15}.   We seek to determine whether viscous dissipation at the ICB can overcome ohmic dissipation in the core and hence be sufficiently high to sustain a dynamo.

This was the strategy followed by \cite{dwyer11}. They estimated the dissipation produced by viscous friction from the differential precession at the CMB as a function of the evolving lunar orbit.  They showed that earlier in the lunar history, this dissipation exceeded by a large amount the power required to maintain the fluid core in an adiabatic state, hence that the remaining power was available to drive a dynamo. The heat dissipated at the CMB is not available to drive thermal convection, as it flows upwards into the mantle or pools at the top of the core.  Hence, a lunar dynamo driven by precession at the CMB invariably depends on whether the vigour and geometry of the core flows forced by the mantle precession can generate and sustain a magnetic field.

In this work, we extend the idea of a dynamo generated by differential precession to the ICB. We investigate whether the viscous friction at the ICB from the differential precession between the inner core and fluid core may have dissipated enough heat in the past to sustain a dynamo.  One key difference is that, in contrast to the heat released at the CMB, that released at the ICB is available to power a thermally driven convective dynamo in the fluid core.  Hence, a dissipation at the ICB higher than the power required to maintain the fluid core in an adiabatic state provides a more robust condition for the presence of a dynamo than the equivalent statement at the CMB.  Core flows forced by a precessing elliptical inner core may further help (or oppose) the generation of a magnetic field, but we do not consider their dynamo capability in our study, nor their influence on a thermally driven dynamo.
 
We build predictions of the differential rotation at both the ICB and CMB using the Cassini state model presented in SD18.  This model allows one to calculate the misalignment between the rotation vectors  of the mantle, inner core and fluid core for a given interior model of the Moon and a set of orbital parameters.   The calculation of the dissipation at the CMB presented in \cite{dwyer11} assumed that the spin vector of the fluid core remained aligned with the ecliptic normal.  But the rotational model of SD18  allows us to calculate more precisely the orientation of the spin vector of the fluid core.   A secondary objective is thus to recalculate the power dissipated at the CMB on the basis of this more complete model.


\section{Theory}

\subsection{Interior models of the Moon}

We follow SD18 and assume a simple model of the lunar interior comprised of four layers made up of a solid inner core, a fluid outer core, a mantle, and a thin crust. The outer radii of each of these layers, in the same sequence, are denoted by $r_s$, $r_f$, $r_m$, and $R$, and their densities, assumed uniform, by $\rho_s$, $\rho_f$, $\rho_m$, and $\rho_c$.  Each layer is triaxial in shape, specified by its polar and equatorial flattenings.  For all interior models in the present study, we use a fixed crustal thickness of $h=R-r_m=38.5$ km with density $\rho_c =2550$ kg m$^{-3}$ \cite[e.g.][]{wieczorek13} and an inner core density fixed at $\rho_s=7700$ kg m$^{-3}$ \cite[][]{matsuyama16}.   To build our interior models, we follow the strategy detailed in SD18: for a given set of $r_s$ and $r_f$, the density of the fluid core and mantle are set by matching the lunar mass $M = ( 4 \pi / 3) \bar{\rho} R^3$, where $\bar{\rho} = 3345.56$ kg m$^{-3}$ is the mean density and $R = 1737.151$ km is the mean radius, and the moment of inertia of the solid Moon $I_{sm} = 0.393112 \cdot M R^2$ \cite[][]{williams14}, comprised here of the mantle and crust.  The polar and equatorial flattenings at each boundary are constrained by matching the degree 2 gravitational potential coefficients $J_2$ and $C_{22}$ as well as the observed surface polar and equatorial flattenings.  We further assume that the ICB and CMB are both at hydrostatic equilibrium with the imposed gravitational potential from the mantle and crust.  

\subsection{Extrapolating the Cassini state and differential rotation in the past}

From the perspective of the rotational dynamics, the mantle and crust are welded together and rotate as a single body, a body which we refer to as the ``mantle" in the context of the lunar rotation.  Hence, the Moon has three independently rotating regions, this ``mantle", the fluid core and the inner core.  Assuming a Cassini state equilibrium, the rotation and figure axes of the mantle and inner core, and the rotation axis of the fluid core can be misaligned from one another, but should all lie in a common plane (the Cassini plane) which also includes the ecliptic and orbit normals.  The Cassini state model developed in SD18 allows one to calculate the mutual orientations of each of these axes.  More specifically, it gives the tilt angles of: the mantle symmetry axis with respect to the ecliptic normal ($\theta_p$); the rotation vector of the mantle ($\theta_m$) and the symmetry axis of the inner core ($\theta_n$), both with respect to the symmetry axis of the mantle; and  the rotation vectors of the fluid core ($\theta_f$) and solid inner core ($\theta_s$), both with respect to the mantle rotation vector. (See Figure 2 of SD18 for a visual representation of each of these angles.)  

For a given interior model, the solution depends on a set of orbital parameters which include the inclination $I$, the eccentricity $e_L$, the precession frequency $\Omega_p$ of the lunar orbit and the sidereal frequency of lunar rotation $\Omega_o$. The ratio of the latter two form the Poincar\'e number, $\delta \omega = \Omega_p / \Omega_o$.  Present-day values for these quantities are $I = {5.145}^\circ$, $e_L=0.0549$, $\Omega_p=2\pi/18.6$ yr$^{-1}$,  $\Omega_o = 2 \pi/27.322$ day$^{-1}$, and $\delta \omega = 4.022 \cdot 10^{-3}$.

Once a solution is obtained, the differential rotation at the CMB and ICB can be deduced.  Viewed by an observer in the mantle frame, the misaligned rotation axis of the fluid core is precessing in a retrograde direction at a frequency of $\omega = \Omega_o + \Omega_p$, so the amplitude and orientation of the differential velocity at the CMB varies with location and time.  Likewise, for the differential velocity at the ICB.  A useful measure of the differential motion is given by the maximum amplitude of the differential angular velocity in the equatorial direction at each of the CMB and ICB.  We denote these as $\Delta \omega_{cmb}$ and $\Delta \omega_{icb}$, respectively, and they are related to $\theta_f$ and $\theta_s$ by

\begin{subequations}
\begin{align}
\Delta \omega_{cmb} & = \Omega_o \big| \sin \theta_f  \big| \, , \label{eq:domcmb} \\
\Delta \omega_{icb} & = \Omega_o  \big| \sin (\theta_f-\theta_s) \big|  \, .\label{eq:domicb}
\end{align}
\end{subequations}
The dissipation at the CMB and ICB can be cast as a function of $\Delta \omega_{cmb}$ and $\Delta \omega_{icb}$, respectively, as we show in the next subsection.   

An order of magnitude for $\Delta \omega_{cmb}$ at present day is readily obtained by assuming a fluid core rotation vector perfectly aligned with the ecliptic normal, and so $\theta_f = -\theta_p = -1.543^\circ$, giving $\Delta \omega_{cmb} = 7.17 \times 10^{-8}$ s$^{-1}$.  Taking a CMB radius of $r_f = 400$ km gives a differential velocity at the CMB of the order of 3 cm s$^{-1}$ and an associated Reynolds number $Re = r_f^2 \,  \Delta \omega_{cmb}/ \nu$ of the order of $10^{11}$ for a kinematic viscosity of $\nu = 10^{-6}$ m$^2$ s$^{-1}$.  Such a large Reynolds number indicates that the viscous friction between the fluid core and mantle should induce turbulent flows.  It is based on this argument  that viscous coupling at the CMB of the Moon is assumed to be in a turbulent regime \cite[][]{yoder81,williams01}.  Although the radius of the ICB is smaller, $\Delta \omega_{icb}$ is typically larger than  $\Delta \omega_{cmb}$ because $| \theta_s | > | \theta_f |$, so the Reynolds number associated with differential precession at the ICB is of similar order and viscous coupling at the ICB is also expected to be in a turbulent regime. With $\Omega_o$ increasing going back in time, and likewise for the magnitudes of $\theta_s$ and $\theta_f$ as we show in our results, it is safe to assume that viscous coupling at both the CMB and ICB has remained in a turbulent regime for the whole of the lunar history.

Each of the orbital parameters $I$, $e_L$, $\Omega_p$ and $\Omega_o$ had different values in the past when the Moon was closer to Earth. To build a history of the differential rotation at the CMB and ICB, we must first determine how these orbital parameters have evolved through time, or as we do here, as a function of the semi-major axis of the lunar orbit $a_L$. For simplicity, we will often refer to $a_L$ as the lunar orbit radius.

Assuming the Moon to be tidally locked into a 1:1 spin orbit resonance, using Kepler's 3rd law, the rotational frequency of the Moon in the past ($\Omega_o(a_L)$) as a function of $a_L$ is given by 

\begin{equation}
\Omega_o(a_L) = \bigg[\frac{a_o}{a_L}\bigg]^{3/2} \Omega_o (a_o) \ ,
\label{eq:orbit}
\end{equation}
where $a_o$ is the present-day semi-major axis equal to $60.3 R_E$, where $R_E$ is the Earth's mean spherical radius.  A numerical integration of the tidal evolution of the Earth-Moon system must be carried out in order to determine how $I$, $e_L$ and $\Omega_p$ have varied as a function of $a_L$.  Examples of such computations can be found in \cite{touma94} and more recently in \cite{cuk16}.  

We restrict our investigation to lunar orbital radii greater than $34 R_E$, hence after the Cassini state transition that occurred at approximately $29 R_E$  \cite[][]{cuk16}.  We take the variation of $\Omega_p$ as a function of $a_L$ presented in Figure 19 of \cite{touma94}.  It is often assumed that changes in $I$ have not been significant after $a_L > 34 R_E$ based on the results of \cite{touma94} (e.g. their Figure 16),   However, \cite{cuk16} have shown that tidal dissipation have lead to substantial changes in $I$, from approximately $18^\circ$ at $a_L = 34 R_E$ to its present-day value of $5.145^\circ$ (see their Figure 4a).  We use the following model for the evolution of $I$, 

\begin{equation}
I = c_1 + c_2 \bigg(\frac{a_o}{a_L}\bigg)^{6}  \, ,
\end{equation} 
with coefficients $c_1 = 4.71976^\circ$ and $c_2 = 0.425237^\circ$; this gives a good approximation to the evolution of $I$ presented in \cite{cuk16}. The eccentricity of the orbit has also varied with $a_L$ although, for simplicity, we assume a fixed eccentricity equal to today's value of $e_L=0.0549$.

An important caveat of our model is that, while we make predictions of the dissipation at the ICB and CMB based on how $\Omega_o$, $\Omega_p$ and $I$ have changed as a function of $a_L$, we do not take into account the feedback that this internal dissipation may have on the evolution of the lunar orbit. Instead, we follow a simplified approach whereby we evaluate {\em a posteriori} whether these predictions are consistent with the lunar orbit history model that we have used.

\subsection{Viscous coupling from turbulent flow at the CMB}

Assuming a turbulent boundary layer, the shear stress acting on the solid boundary can be written as

\begin{equation}
\bm{\tau}= f \rho_f \left|\bf{u}\right| \bf{u} \ ,
\label{eq:turbulent_stress}
\end{equation} 
where $\rho_f$ is the density of the fluid core, $\bf{u}$ is the flow velocity outside the boundary layer, and $f$ is a dimensionless coefficient of friction which depends, among other things, on surface roughness.  Integrated over the CMB, the amplitude of the torque at the CMB resulting from this turbulent viscous shear stress can be written in the form

\begin{equation}
{\Gamma_{cmb}} = f_{cmb} \, \bar{C}_f \, \big| {\Delta \omega_{cmb}} \big|^2   \, ,\label{eq:tcmb1}
\end{equation}
where $\Delta \omega_{cmb}$ is given by Equation (\ref{eq:domcmb}) and where $\bar{C}_f = (8 \pi /15) \rho_f \,r_f^5$ is the mean moment of inertia of an entirely fluid core.  $f_{cmb}$ is a coefficient of friction, though different in numerical value from $f$ as it takes into account the integration of the stress over the whole surface of the CMB.  Constraints on the amplitude of the viscous friction at the CMB of the Moon at present-day can be derived from LLR observations \cite[][]{williams01,williams14,williams15}, and provide an estimate of $f_{cmb}$. The rotational model of the Moon used to fit LLR data consists of a rigid mantle and a fluid core (it does not include an inner core). Viscous dissipation is incorporated into the model by prescribing a viscous torque on the mantle in the form

\begin{equation}
{\Gamma_{cmb}}={\cal K} \cdot \Delta \omega_{cmb} \, ,\label{eq:tcmb2}
\end{equation} 
where $\mathcal K$ is a coupling coefficient.  A recent estimate of ${\cal K}$ is \cite[e.g.][]{williams15}

\begin{equation}
\frac{\cal K}{\bar{C}} =  (1.41 \pm 0.34) \times 10^{-8} \; \mbox{days}^{-1} =(1.63 \pm 0.39) \times 10^{-13} \; \mbox{s}^{-1} \, ,
\label{eq:kc}
\end{equation}
where $\bar{C} = (8 \pi /15) \bar{\rho} \, R^5$ is the mean moment of inertia of the whole Moon. Equating Equations (\ref{eq:tcmb1}) and (\ref{eq:tcmb2}), an estimate of $f_{cmb}$ at present-day is then given by

\begin{equation}
 f_{cmb} = \left( \frac{{\cal K}}{\bar{C}}\right) \left( \frac{\bar{C}}{\bar{C}_f}\right) \frac{1}{\big| \Delta \omega_{cmb}  \big|_{today}} \, , \label{eq:fcmb}
\end{equation}
where the subscript {\em today} emphasizes that it is based on the present-day value of the differential rotation at the CMB.  For a given interior density model of the Moon, we can calculate the ratio $\bar{C}_f/\bar{C}$ and determine ${\big| \Delta \omega_{cmb}  \big|_{today}}$ using the Cassini state model of SD18.  Hence, we can readily calculate $f_{cmb}$. To present an estimate, using $\theta_f=-1.6^\circ$ [e.g. SD18],  $\bar{C}_f/\bar{C} = 7 \cdot 10^{-4}$ \cite[e.g.][]{williams14} and ${\cal K}/{\bar{C}}$ from Equation (\ref{eq:kc}) gives $f_{cmb} =0.0314$.

Viscous dissipation at the CMB (${Q}_{cmb}$) can be calculated as the product of the torque ${\Gamma_{cmb}}$ and the angular velocity difference $\Delta \omega_{cmb}$,

\begin{equation}
\label{eq:dissipation_general}
{Q}_{cmb}={\Gamma_{cmb}} \cdot \Delta \omega_{cmb}  =f_{cmb} \, \bar{C}_f \, \big| {\Delta \omega_{cmb}} \big|^3  \, .
\end{equation}
Based on the estimate of $f_{cmb}$ at present-day from Equation (\ref{eq:fcmb}), and assuming that $f_{cmb}$ has remained constant, the dissipation in the past,  when $\Omega_o$ and $\theta_f$ were both different, can be obtained from 

\begin{equation}
{Q}_{cmb}  = \bar{C} \left( \frac{{\cal K}}{\bar{C}}\right)  \frac{\big| \Omega_o \sin \theta_f \big|_{past}^3}{\big| \Omega_o \sin \theta_f \big|_{today}} \,  .\label{eq:qcmb}
\end{equation}

A more proper evaluation of ${Q}_{cmb}$ should take into account the fact that $f_{cmb}$ depends on $\Omega_o$ \cite[e.g.][]{cebron19}.  However, the dependence is weak, and since our primary objective is to derive an order of magnitude estimate of how ${Q}_{cmb}$ has evolved, we neglect this effect here.

It is important to emphasize that the estimate of ${\cal K}$ obtained from LLR observations is based on a rotational model of the Moon that does not include an inner core.  If an inner core is present, the coefficient ${\cal K}$ in this rotational model captures the combined effect of friction at both the CMB and ICB \cite[][]{williams09}.  Moreover, the manner in which the delayed tidal response of the Moon varies with different forcing frequencies -- ultimately the method by which an estimate of ${\cal K}$ separate from tidal deformation is obtained \cite[e.g.][]{williams01,williams15} -- can also be different if an inner core is present.  Additionally, viscous deformation within the inner core may also contribute to a part of the observed dissipation in lunar rotational energy.   These caveats are mentioned to keep the reader alert to the fact that the estimate of the present-day viscous dissipation at the CMB (and at the ICB) remain not well constrained by observations.  Since our results are ultimately tied to value of ${\cal K}$ given by Equation \ref{eq:kc}, they must be interpreted as order of magnitude estimates.

\subsection{Viscous coupling from turbulent flow at the ICB}

By analogy with the torque at the CMB, the amplitude of the torque at the ICB resulting from turbulent viscous shear stress is given by

\begin{equation}
\Gamma_{icb} =  f_{icb} \, \bar{C}_s \,\left( \frac{\rho_f}{\rho_s}\right)  \big| \Delta \omega_{icb}  \big|^2 \, ,
\end{equation}
where $\bar{C}_s = (8 \pi /15) \rho_s \,r_s^5$ is the mean moment of inertia of the inner core, the ratio $\rho_f/\rho_s$ accounts for the fact that it is the density of the fluid core which is involved in the shear stress, and $f_{icb}$ is a friction coefficient for the ICB. 

 It is not possible to get an independent estimate of $f_{icb}$ based on LLR observations. To move forward and build an estimate of the dissipation at the ICB, we must make an assumption on $f_{icb}$ and we simply assume that it is equal to $f_{cmb}$.  There is no reason a priori why this should be the case, but this is the simplest assumption one can make.  All our results depend on this assumption so they must be viewed with this caveat in mind.

The viscous dissipation at the ICB is

\begin{equation}
\label{eq:dissipation_general}
{Q}_{icb}={\Gamma_{icb}} \cdot \Delta \omega_{icb}  =f_{icb} \, \bar{C}_s \,\left( \frac{\rho_f}{\rho_s}\right) \big| {\Delta \omega_{icb}} \big|^3  \, .
\end{equation}
Setting $f_{icb}=f_{cmb}$ and using $f_{cmb}$ as prescribed by Equation (\ref{eq:fcmb}), the dissipation at the ICB in the past is given by 

\begin{equation}
{Q}_{icb} =\bar{C} \left( \frac{r_s}{r_f} \right)^5 \left( \frac{{\cal K}}{\bar{C}}\right)  \frac{\big| \Omega_o \sin ( \theta_f -\theta_s) \big|_{past}^3}{\big| \Omega_o \sin \theta_f \big|_{today}} \, .\label{eq:qicb}
\end{equation}

\subsection{Magnetic field strength from dissipation at the CMB and ICB}

To convert dissipation at the CMB into magnetic field intensity at the lunar surface, \cite{dwyer11} used a scaling law derived in \cite{christensen09}, based on numerical dynamo models powered by convection. This scaling may not be entirely suitable for a dynamo generated by mechanical stirring at the CMB, but no equivalent scaling law is available yet for precessional dynamos.  We thus proceed similarly here.  We use the notation ${\cal B}_{(cmb)}$ to denote the amplitude of the magnetic field at the lunar surface resulting from mechanical forcing due to precession at the CMB.  The relationship between ${\cal B}_{(cmb)}$ (in units of $\mu$T) and the dissipation available to power the dynamo ($Q_{dyn}$) used in \cite{dwyer11} is

\begin{equation}
{\cal B}_{(cmb)} \approx 6 \,  d \, \bigg(\frac{r_f}{r_{fo}} \bigg)^{3} \bigg(\frac{Q_{cmb}^{dyn}}{\bar{Q} }\bigg)^{1/3} \ ,
\label{eq:Bcmb}
\end{equation}
where $\bar{Q} = 3 \times 10^{11}$ W and $d$ is the ratio of the dipolar magnetic field to the total field at the CMB.  For simplicity, we set $d$ equal to 1, which represents an upper bound for ${\cal B}_{(cmb)}$.  The factor ($r_f/r_{fo}$) takes into account a different choice of core radius than the reference $r_{fo}=350$ km used by \cite{dwyer11}.  $Q_{cmb}^{dyn}$ is the dissipation readily available to power the dynamo and in \cite{dwyer11} it was taken as 

\begin{equation}
Q_{cmb}^{dyn}= Q_{cmb}-Q_{th} \, ,
\end{equation}
where $Q_{cmb}$ is given by Equation (\ref{eq:qcmb}) and $Q_{th}$ is a threshold value below which no dynamo can exists.  $Q_{th}$ was taken in \cite{dwyer11} as the adiabatic heat flow at the CMB, estimated at $4.7 \times 10^9$ W, which represents the minimum heat flow out of the core required in order to sustain a dynamo driven by thermal convection alone \cite[e.g.][]{nimmo15}.  While this may be a valid threshold for convective dynamos, it is less clear that it applies for a mechanical dynamo driven by precession.  Even if no heat flow escapes the core, and the latter is thermally stratified, tidal instabilities can still be generated  \cite[e.g.][]{cebron10,vidal18,vidal19} and the additional mechanical forcing at the boundary from the precessing mantle may be capable to drive a dynamo.  Nevertheless, there must be a threshold value otherwise a precession dynamo would still be operating in the Moon today.  For simplicity, and in the absence of a different estimate, we also use $Q_{th} = 4.7 \times 10^9$ W.

We use the same scaling law given by Equation (\ref{eq:Bcmb}) to determine the amplitude of the magnetic field at the lunar surface resulting from a precession driven dynamo at the ICB, which we denote by ${\cal B}_{(icb)}$, and is given by, in units of $\mu$T, 

\begin{equation}
{\cal B}_{(icb)} \approx 6\, d \, \bigg(\frac{r_f}{r_{fo}} \bigg)^{3}  \bigg(\frac{Q_{icb}^{dyn}}{\bar{Q} }\bigg)^{1/3} \ ,
\label{eq:Bicb}
\end{equation}
where again we set $d$ equal to 1, and with

\begin{equation}
Q_{icb}^{dyn}= Q_{icb}-Q_{th} \, ,
\end{equation}
where $Q_{icb}$ is given by Equation (\ref{eq:qicb}).  The adiabatic heat flow at the ICB is different than at the CMB, but the criteria to maintain a dynamo driven by thermal convection in the fluid core remains tied to the total heat flow escaping the core, so we take again $Q_{th} = 4.7 \times 10^9$ W.  It should be noted that, while heat dissipated at the CMB is not available to drive convection, as it flows upwards into the mantle or pools at the top of the core, heat generated by friction at the ICB is available to power a thermally driven convective dynamo. Hence, the scaling law of Equation (\ref{eq:Bicb}) may be more justified for a dynamo thermally driven by precession at the ICB.  We note that it is likely a lower bound since, in addition to the thermal energy available to drive convective flows, the precession motion of an elliptical inner core also generates flow by mechanical stirring.  If such flows lead to global instabilities and large scale eddies \cite[e.g.][]{lin16}, they may further contribute to (though they might also suppress) dynamo action.  In addition to the uncertainty in estimating $Q_{icb}$, as detailed in the previous section, this further adds to the uncertainty of estimating  ${\cal B}_{(icb)}$.  Hence, our calculation of ${\cal B}_{(icb)}$ from Equation (\ref{eq:Bicb}) must be viewed as an order of magnitude estimate.


\section{Results}

\subsection{Evolution of the Cassini state}

\begin{figure}[!htbp]
\centering
\includegraphics[width=9cm]{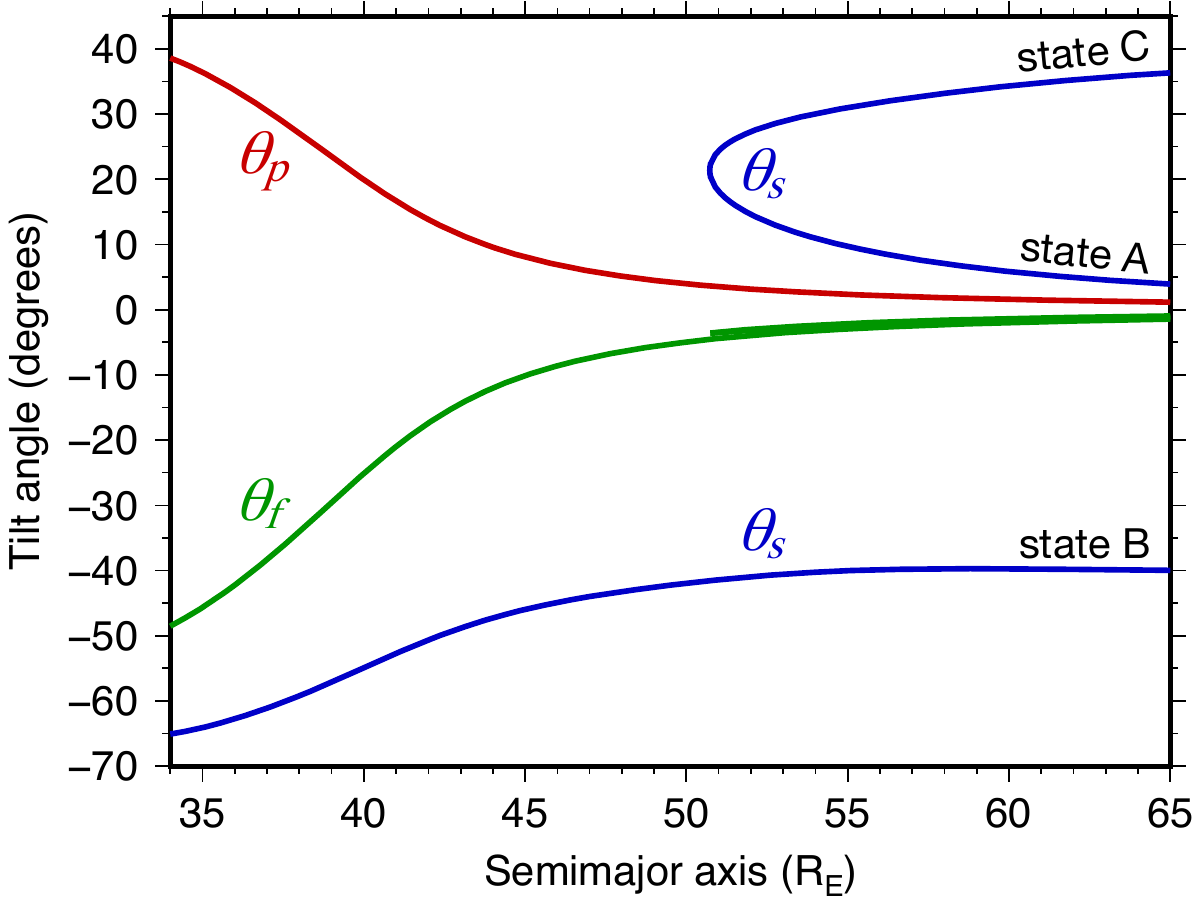}
\caption{\label{fig:theta_vs_aL} Evolution of the Cassini state of a specific Moon model ($r_f=350$ km, $r_s=250$ km) as a function of lunar orbital radius in units of Earth radii ($R_E$).  Shown are the tilt angles of: the mantle symmetry axis with respect to the ecliptic normal ($\theta_p$, red);  the rotation vectors of the fluid core ($\theta_f$, green) and solid inner core ($\theta_s$, blue), both with respect to the mantle rotation vector. States A, B and C refer to the different Cassini states of the inner core. Today corresponds to $60.3 R_E$.}
\end{figure}

We first show an example of how the Cassini state equilibrium of the different interior regions of the Moon changes as a function of orbital radius, and thus how the differential velocity at both the CMB and ICB may evolve.  Figure \ref{fig:theta_vs_aL} shows the evolution of $\theta_p$, $\theta_f$ and $\theta_s$ as a function of the lunar orbit radius for a particular Moon model with a fluid outer core radius of $r_f = 350$ km and a solid inner core radius of $r_s=250$ km. When the Moon was at $34 R_E$, $\theta_p$ was close to ${40}^\circ$, consistent with the results shown in \cite{ward75}. The magnitude of the variation of $\theta_f$ is slightly larger and tracks the changes in $\theta_p$, though with the reverse sign. Recall that $\theta_f$ is measured with respect to the spin vector of the mantle $\theta_m$, and the latter always remains closely aligned with the symmetry axis of the mantle. Hence, Figure \ref{fig:theta_vs_aL} shows that, for $a_L > 34 R_E$, the rotation axis of the fluid core always remain closer to an alignment with the ecliptic normal than to an alignment with the mantle.  However, it is important to point out that the spin axis of the fluid core is always misaligned with the ecliptic normal, in the opposite direction than the mantle tilt, and that the offset gets larger the smaller $a_L$ is.  This is consistent with the recent results of \cite{cuk19}.  For $a_L = 34 R_E$, the offset is as large as ${10}^\circ$.  The reason why the rotation vector of the fluid core never lines up closely with the mantle symmetry axis is because, for all values of $a_L$ in Figure \ref{fig:theta_vs_aL},  the FCN frequency is always much smaller than the precession frequency $\Omega_p$ \cite[e.g.][]{meyer11}. 

\begin{figure}[!htbp]
\centering
\includegraphics[width=9cm]{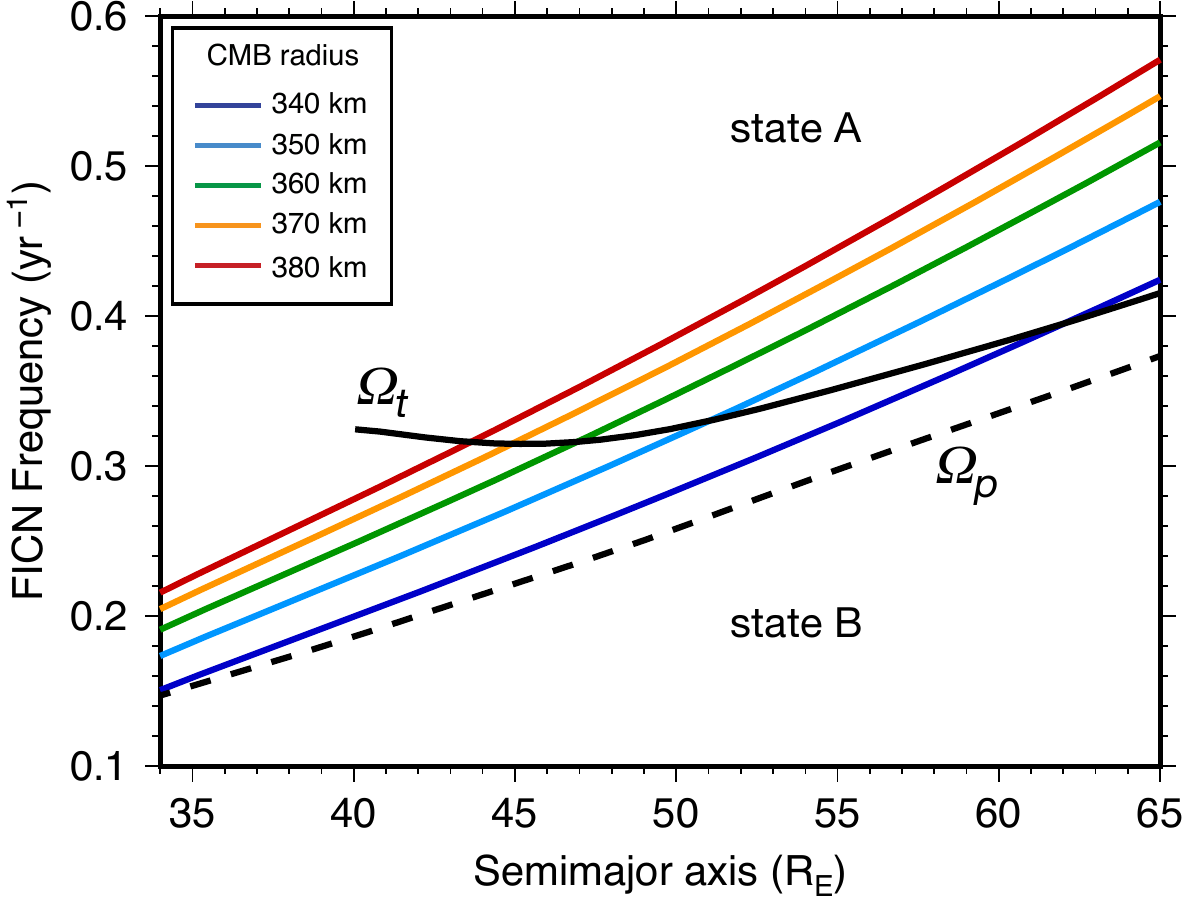}
\caption{\label{fig:ficn_vs_aL} Evolution of the FICN frequency $\Omega_{ficn}$ as a function of lunar orbital radius in units of Earth radii ($R_E$), for different choices of outer core radii, and for a solid inner core radius of 250 km.  The precession frequency $\Omega_p$ is indicated by the dashed line.  The Cassini state transition frequency $\Omega_t$ of the inner core is indicated by the black solid curve. The inner core occupies state B when $\Omega_{ficn}<\Omega_t$, and state A when $\Omega_{ficn}>\Omega_t$.  Today corresponds to $60.3 R_E$.}
\end{figure}

For all values of $a_L$ on Figure \ref{fig:theta_vs_aL}, the spin axis of the inner core $\theta_s$ (which remains closely aligned with the symmetry axis of the inner core $\theta_n$) is significantly offset from the mantle symmetry axis.  This is because, for all $a_L$,  the FICN frequency remains sufficiently close to the forcing (precession) frequency (see Figure  \ref{fig:ficn_vs_aL}) and a large tilt of the inner core results from resonant amplification.   Moreover, Figure \ref{fig:theta_vs_aL} shows that the solid inner core can occupy different Cassini states.  The states labelled A, B and C follow the convention introduced in SD18.   For this specific Moon model, only state B was possible for $a_L < 51 R_E$, but all three states are possible solutions for $a_L > 51 R_E$.  The orientation of the spin vector of the fluid core is slightly different in each of these three Cassini states (see SD18) and for a large inner core, as is the case here, the shift in $\theta_f$ is sufficiently large than it can be seen in Figure  \ref{fig:theta_vs_aL}.

Assuming that the lowest energy state is favoured (i.e. the state with the smallest inner core tilt), Figure  \ref{fig:theta_vs_aL} shows that, as the Moon moved away from the Earth, it is possible that a transition in the Cassini state of the inner core from state B to state A may have occurred.  Whether such a transition did take place depends on the core density model.   As shown in Figure 4 of SD18, the Cassini state that the inner core occupies today is determined by how the FICN frequency compares with the precession frequency $\Omega_p$.  The FICN frequency is retrograde (as is $\Omega_p$) and we use here the notation $\Omega_{ficn}$ to denote its amplitude, as it is seen by an observer in a space-fixed frame.  The transition between states B and A does not occur exactly at $\Omega_{ficn}=\Omega_p$, but instead at $\Omega_{ficn}=\Omega_p +\delta \Omega$, where $\delta \Omega$ is a correction that involves the tilt angles $\theta_n$ and $\theta_p$. Denoting this ``transition'' frequency by $\Omega_t = \Omega_p + \delta \Omega$, if ${\Omega}_{ficn} < \Omega_t$, the inner core occupies state B; if ${\Omega}_{ficn} > \Omega_t$, it occupies state A. For the present-day Moon, $\Omega_t = 2 \pi/16.4$ yr$^{-1}$ (SD18).

This rule applies at any moment in the lunar history, so a Cassini state transition from states B to A implies an intersection  between ${\Omega}_{ficn}$ and $\Omega_t$ as they both evolve.    Figure \ref{fig:ficn_vs_aL} shows how $\Omega_{ficn}$ changes as a function of $a_L$ for a Moon model with $r_s=250$ km and different choices of fluid outer core radii, as well as how the transition frequency $\Omega_t$ varies as a function of $a_L$.   The lunar model with $r_f=350$ km whose tilt angles' evolution are shown in Figure  \ref{fig:theta_vs_aL} corresponds to the light blue curve in Figure \ref{fig:ficn_vs_aL}: for this model, ${\Omega}_{ficn}$ intersects $\Omega_t$ at $a_L = 51 R_E$, which marks the point at which the Cassini state transition from state B to state A occurs.  

To understand how $\Omega_{ficn}$ changes with $a_L$, to a good approximation it is given by Equation 24 of SD18 multiplied by $\Omega_o$ (and with a reversed sign, since we define here a retrograde frequency as being positive),

\begin{equation}
\Omega_{ficn}  = - \Omega_o e_s \alpha_1 +  \Omega_o e_s \alpha_g \big(1- \alpha_1 \big) + \frac{3}{2}  \Omega_o \beta_s \big(1- \alpha_1 \big) \big(\cos^2 I - \sin^2 I \big)  \, , \label{eq:omficn}
\end{equation}
where we have assumed $e_L=0$ to simplify and where $\alpha_1=\rho_f/ \rho_s$, $e_s=(C_s - \bar{A}_s)/\bar{A}_s$, $\beta_s=(C_s - {A}_s)/\bar{A}_s$, with $C_s$, ${A}_s$ and $\bar{A}_s$ being respectively the polar, minimum and mean equatorial moments of inertia of the inner core.  The parameter $\alpha_g$ is given by Equation (18) of SD18 and represents the ratio of the gravitational to the centrifugal (or inertial) pressure torque exerted on the inner core. For a non-evolving lunar density structure, $\alpha_g$ was smaller in the past because the rotation rate $\Omega_o$ was higher and thus the centrifugal torque was relatively more important.  In fact,  $\alpha_g$ is proportional to $\Omega_o^{-2}$. The second term on the right-hand side of Equation (\ref{eq:omficn}), which features $\alpha_g$, dominates the two other terms.  Hence, $\Omega_{ficn}$ is proportional to $\Omega_o \alpha_g$, and thus is inversely  proportional to $\Omega_o$.  Since $\Omega_o$ decreases with $a_L$, $\Omega_{ficn}$ increases with $a_L$ as seen in Figure \ref{fig:ficn_vs_aL}.  $\alpha_g$ depends on the density contrast at the ICB, so the $\Omega_{ficn}$ curves for different $r_f$ shown in Figure \ref{fig:ficn_vs_aL} would be displaced with a different choice of inner core radius.  However, note that the transition frequency is to first order independent of the density structure in the lunar core; for any choice of core density model, the Cassini state transition is determined by when the $\Omega_{ficn}$ curve for this particular model intersects the $\Omega_{t}$ curve that is shown in Figure \ref{fig:ficn_vs_aL}.  

A Cassini state transition has important implications for a dynamo driven by inner core precession.  First, the change from states B to A is not instantaneous, and the large change in inner core tilt that it involves (from approximately ${-42}^\circ$ to ${+21}^\circ$ on Figure \ref{fig:theta_vs_aL}) should generate instabilities and flows in the fluid core.  For a relatively short period of time, these flows may be capable of generating a dynamo.  After the transition, and once instabilities in the fluid core have attenuated, the differential velocity at the ICB associated with state A is much smaller, implying a sudden drop in the power dissipated at the ICB and the potential for dynamo action.

\subsection{Power dissipation at the CMB and ICB versus lunar orbit radius}

\begin{figure}[!htbp]
\centering
\includegraphics[width=9cm]{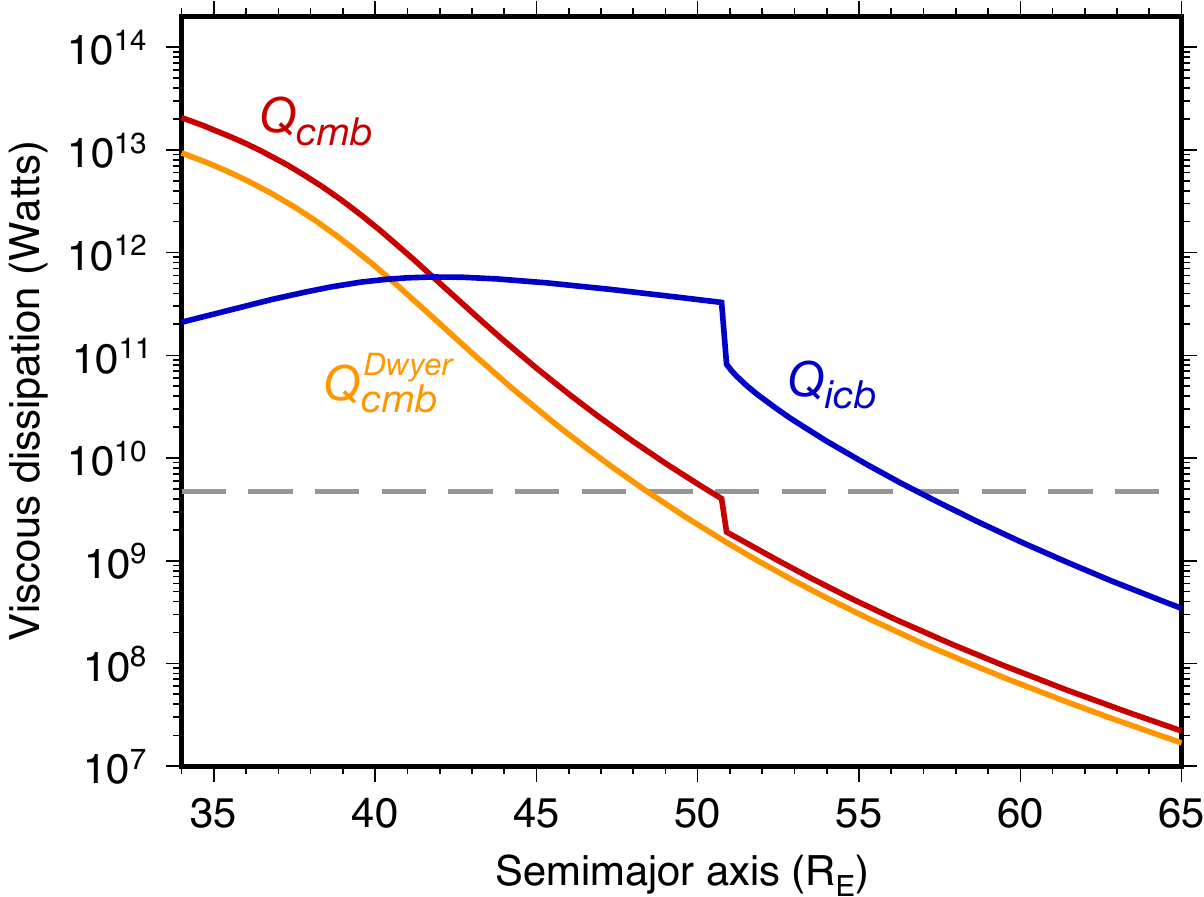}
\caption{ \label{fig:Qicb1} Power available to drive a dynamo estimated from viscous dissipation at the CMB ($Q_{cmb}$, red) and at the ICB ($Q_{icb}$, blue) as a function of lunar orbit radius in units of Earth radii ($R_E$). The dissipation at the CMB from the model presented in \cite{dwyer11} is shown in orange ($Q_{cmb}^{Dwyer}$).  The evolution of the differential velocities at the CMB and ICB correspond to those shown on Figure \ref{fig:theta_vs_aL}.  The power threshold to sustain a dynamo ($Q_{th}=4.7\times10^9$ W) is indicated by the grey horizontal dashed line. Today corresponds to $60.3 R_E$.}
\end{figure}

The dissipation at the CMB and ICB can be expressed as a function of the lunar orbit by Equations (\ref{eq:qcmb}) and (\ref{eq:qicb}), respectively.  Figure \ref{fig:Qicb1} shows how the viscous dissipation at both the CMB and ICB vary as a function of $a_L$ for the same lunar model with $r_f = 350$ km and $r_s = 250$ km, whose tilt angles' evolution is shown in Figure \ref{fig:theta_vs_aL}.   For comparison, we also show  on Figure \ref{fig:Qicb1} the dissipation at the CMB estimated by  \cite{dwyer11}, using the same core radius of $r_f = 350$ km, and computed from

\begin{equation}
Q_{cmb}^{Dwyer} \approx 3 \times 10^{20} W \times \frac{\sin{^3\theta_p}}{(a_L(t)/R_E)^{(9/2)}} \, .
\label{fig:dwyer_dissipation}
\end{equation}
Our estimate of $Q_{cmb}$ differs from that of \cite{dwyer11} for two reasons.  First, Equation  (\ref{fig:dwyer_dissipation}) is based on a dissipation at present day of $Q_{cmb} \approx (5.8 \pm 1.3) \times 10^7$ W which is itself derived from a viscous coupling coefficient of ${\mathcal K}/\bar{C} \approx 1.122 \pm 0.257 \times 10^{-8}$ day$^{-1}$ estimated in \cite{williams01}. Using instead the updated value of ${\mathcal K}/\bar{C}$ given in Equation (\ref{eq:kc}) gives a larger present-day dissipation of $7.3 \pm 1.8 \times 10^7$ W and our higher estimates of $Q_{cmb}$ are in part due to this.  Second, the reconstruction of $Q_{cmb}$ in Equation (\ref{fig:dwyer_dissipation}) makes the implicit assumption that the spin axis of the fluid core has remained aligned with the ecliptic normal.  As we have shown above (and in DW16 and SD18), this is incorrect: the spin axis of the fluid core is offset from the ecliptic normal, in the reverse direction than the mantle offset, resulting in a larger angle of offset between the rotation vectors of the fluid core and mantle.  This difference is larger the further we go back in time and this also contributes to make our estimates of $Q_{cmb}$ larger.  Note that as a result of the Cassini state transition associated with the inner core at $a_L = 51 R_E$, there is a drop in the differential velocity at the CMB, and thus a drop in $Q_{cmb}$.  The dashed grey line on Figure \ref{fig:Qicb1} corresponds to the power required to maintain an adiabat in the fluid core.  \cite{dwyer11} used this value as the threshold power for dynamo action, $Q_{th}$.   We have already pointed out that this may not be an appropriate lower bound for a mechanically forced dynamo, but if we adopt this specific choice, the intersection between $Q_{cmb}$ and $Q_{th}$ occurs at $a_L \approx 50.4 R_E$, slightly before the inner core Cassini state transition, at which point the dynamo from $Q_{cmb}$ ceases.

We also show on Figure \ref{fig:Qicb1} the power dissipated at the ICB, $Q_{icb}$. Not only is $Q_{icb}$ higher than $Q_{cmb}$ for $a_L > 42$ $R_E$, it remains above the dynamo threshold for a much longer period of time, therefore allowing for a dynamo that may have persisted to more recent epochs. The sudden drop in $Q_{icb}$ at $a_L \approx 51$ $R_E$ is due to the transition of the Cassini state of the inner core, from state B to state A, the latter featuring a smaller differential rotation at the ICB. Note that a large scale flow reorganization in the core may accompany this Cassini transition, which would lead to a spike in $Q_{icb}$ (and also $Q_{cmb}$), before settling to the lower energy state. However, we cannot model this with our idealized Cassini state equilibrium model.

Different combinations of $r_s$ and $r_f$, and thus, in general, different interior density models of the Moon, lead to changes in the predicted variations of $\theta_f$ and $\theta_s$ with $a_L$.  Changes in $\theta_f$ remain modest for different interior models, with differences not exceeding 1 degree from the evolution scenario shown in Figure  \ref{fig:theta_vs_aL}.  These lead to modifications in the predicted amplitude of $Q_{cmb}$ versus $a_L$ depicted in Figure \ref{fig:Qicb1}, but not by more than approximately 10\%.   

\begin{figure}[!htbp]
\centering
\includegraphics[height=5.5cm]{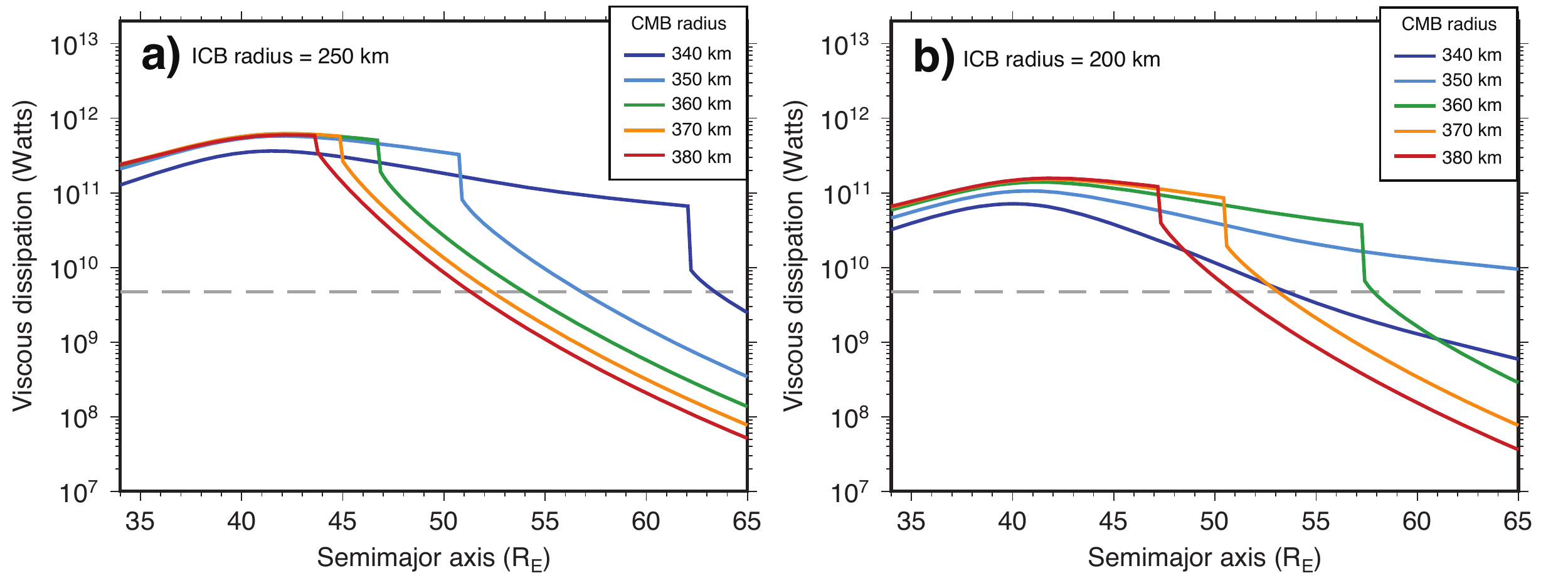}
\caption{\label{fig:Qicb2} Power available to drive a dynamo estimated from viscous dissipation at the ICB ($Q_{icb}$) as a function of lunar orbit radius in units of Earth radii ($R_E$), for different choices of outer core radii, and for a solid inner core radius of (a) 250 km and (b) 200 km.  The power threshold to sustain a dynamo ($Q_{th}=4.7\times10^9$ W) is indicated by the grey horizontal dashed line. Today corresponds to $60.3 R_E$.}
\end{figure}

In contrast, the way in which $\theta_s$ (and thus  $Q_{icb}$) vary with $a_L$ is highly sensitive to the choice of lunar interior density model.   To illustrate this, Figure \ref{fig:Qicb2} shows how $Q_{icb}$ vary with $a_L$ for five different choices of outer radii (340, 350, 360, 370 and 380 km) and two different choices of inner core radii (250 km for panel a; and 200 km for panel b).  Since $Q_{icb}$ is proportional to $r_s^5$ (see Equation \ref{eq:qicb}), the size of the inner core is of crucial importance for the power available to drive a dynamo, and by reducing $r_s$ from 250 to 200 km,  the maximum $Q_{icb}$ has dropped by approximately a factor of 4.  In addition, different interior density models lead to different histories of $Q_{icb}$.  This is because the different combinations of $r_s$ and $r_f$ imply a different fluid core density in each of these interior models in order to match the lunar mass.  In turn, the different density structure affects the FICN frequency $\Omega_{ficn}$ of the lunar model.  Since the tilt angle of the inner core is determined by the relative difference between $\Omega_{ficn}$ and the forcing frequency $\Omega_p$, the differential velocity at the ICB, and thus $Q_{icb}$, shows very different histories for different lunar models.  Moreover, the point in time at which $\Omega_{ficn}$ intersects the transition frequency $\Omega_t$ can be substantially changed, and hence so does the timing of a Cassini transition.   For a given inner core size, the smaller the fluid outer core radius, the later in lunar history the transition occurs (see Figure \ref{fig:ficn_vs_aL}). For the model with $r_s = 250$ km and $r_f = 340$ km (Figure \ref{fig:Qicb2}a), and for the models with $r_s = 200$ km and $r_f = 340$ and $350$ km  (Figure \ref{fig:Qicb2}b), the transition has not yet occurred. 

\subsection{Paleomagnetic intensity from power dissipation}

\begin{figure}[!htbp]
\centering
\includegraphics[height=5.5cm]{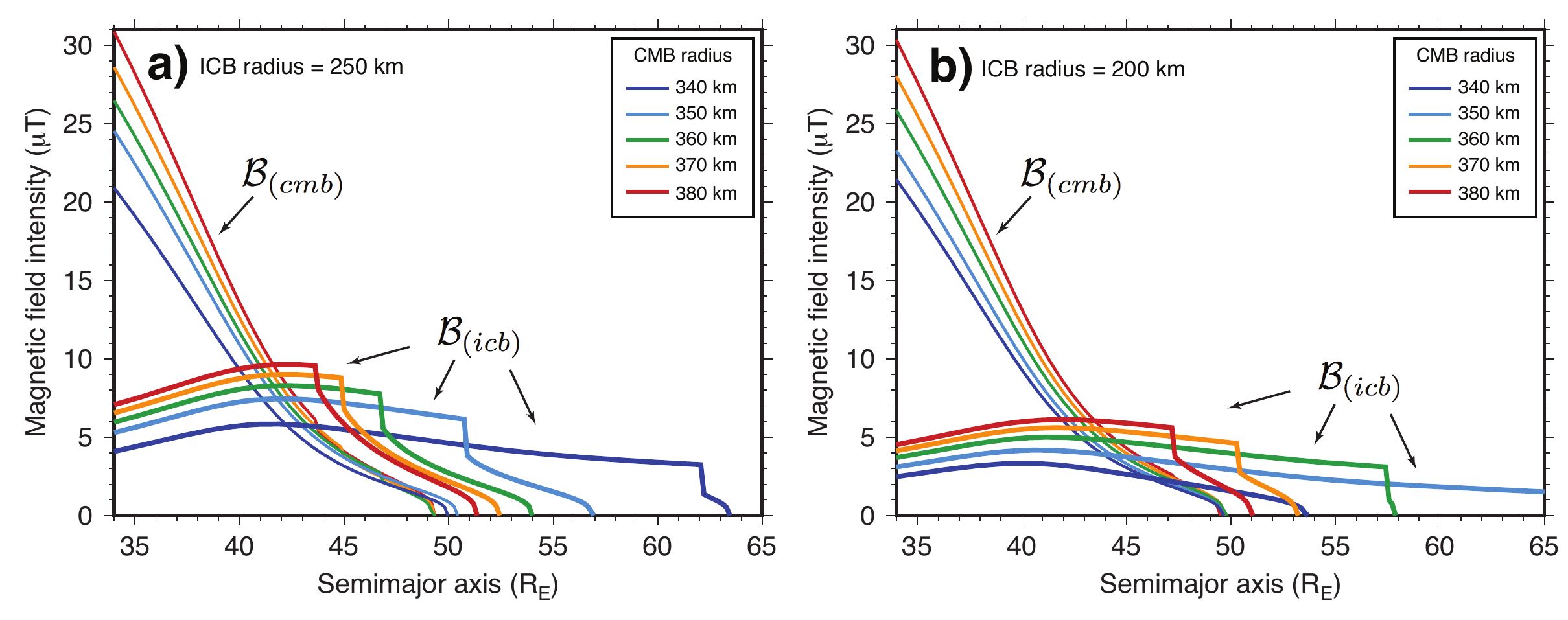}
\caption{\label{fig:Bicb} Paleomagnetic field intensities at the Moon's surface from viscous dissipation at the CMB (${\cal B}_{(cmb)}$) and ICB (${\cal B}_{(icb)}$) as a function of lunar orbit radius in units of Earth radii ($R_E$), for different choices of outer core radii, and for a solid inner core radius of (a) 250 km and (b) 200 km. Today corresponds to $60.3 R_E$.}
\end{figure}

For the same set of models as in Figure \ref{fig:Qicb2}, Figure \ref{fig:Bicb} shows the magnetic field strengths ${\cal B}_{(cmb)}$ and ${\cal B}_{(icb)}$ at the lunar surface predicted from dissipation at the CMB and at the ICB, respectively, and as a function of $a_L$.  Dissipation at the CMB leads to a ${\cal B}_{(cmb)}$ field in the range of 20-30 $\mu$T for $a_L=34 R_E$, although its amplitude drops rapidly with increasing $a_L$.  Note that the change in ${\cal B}_{(cmb)}$ between the different models is dominantly caused by the factor of $r_f$ in Equation (\ref{eq:Bcmb}).  For all cases shown in Figure \ref{fig:Bicb}, $Q_{cmb}$ drops below $Q_{th}$ between $a_L=49 R_E$ and $50.4 R_E$, marking the point at which the dynamo from dissipation at the CMB ceases.  Note also that when a transition in the Cassini state associated with the inner core occurs, there is a drop in $Q_{cmb}$, leading to a smaller  ${\cal B}_{(cmb)}$ and an earlier dynamo shutoff; this is most evidently seen in Figure \ref{fig:Bicb}a for the largest inner core.  

Dissipation at the ICB leads to a surface field amplitude ${\cal B}_{(icb)}$ which can be as high as 10 $\mu$T.  We recall that our estimates of the magnetic field strength are based on $d=1$ in Equation (\ref{eq:Bicb}), in other words that all the magnetic energy is assumed to be in the dipole part.  Since ${\cal B}_{(icb)}$ scales linearly with $d$, the predictions on Figure \ref{fig:Bicb} would decrease in proportion with a smaller choice for $d$. However, we also recall that the scaling law that we use for ${\cal B}_{(icb)}$ is based on convective dynamos, and the flows induced by the precession of an elliptically shaped inner core may further contribute to dynamo action and thus lead to an increased ${\cal B}_{(icb)}$.  For all cases shown in Figure \ref{fig:Bicb}, the dynamo from $Q_{icb}$ shuts off after that from $Q_{cmb}$.  While not as strong earlier in lunar history, the dynamo from precession at the ICB may have persisted for a much longer period.  

The dissipation and magnetic field strength at the ICB shown in Figures \ref{fig:Qicb2} and \ref{fig:Bicb} are for relatively large inner core radii of 250 and 200 km. Since $Q_{icb}$ scales with $r_s^5$ (see Equation \ref{eq:qicb}) and thus $B_{(icb)}$ scales with $r_s^{5/3}$ (from Equation \ref{eq:Bicb}), it is clear that smaller $r_s$ would yield smaller $Q_{icb}$ and $B_{(icb)}$.  There is a critical inner core size below which $Q_{icb}$ is smaller than the dynamo threshold $Q_{th}$ even at small $a_L$.  This is explored in the next subsection.

Figures \ref{fig:Qicb2} and \ref{fig:Bicb} illustrate how the histories of $Q_{icb}$ and ${\cal B}_{(icb)}$ are sensitive to the density structure of the lunar core.  The evolution scenarios presented in these figures only take into account the changes in the lunar orbit through its effect on $\Omega_o$, $\Omega_p$ and $I$, and assume a non-changing core density structure.  However, the presence of an inner core is due to growth from crystallization, and hence the radius of the inner core should increase as a function of time. As the inner core grows, assuming it is composed primarily of Fe, the proportion of lighter elements in the fluid core increases.  The changing density contrast at the ICB implies a change in FICN frequency which can affect the evolution scenarios presented above.

The precise history of the inner core growth depends among other things on the initial composition of the fluid core and on the evolution of the heat flux at the CMB \cite[e.g.][]{laneuville14} which are not well known.  To add to this difficulty, attaching a precise time-history to the lunar orbit is challenging.  Instead of presenting a possible evolution scenario that takes into account inner core growth, here we simply mention how our above results would need to be adapted.  Since $Q_{icb}$ is highly sensitive to $r_s$, taking inner core growth into account would make $Q_{icb}$ in Figure \ref{fig:Qicb2} weaker at the earliest $a_L$ that we have considered.  Not only ${\cal B}_{(icb)}$ would be reduced, since $Q_{icb}$ may fall below $Q_{th}$ for smaller $a_L$, a dynamo powered by differential precession at the ICB may only have started later in lunar history.  


\subsection{Power dissipation at the ICB and magnetic field strength for a suite of interior Moon models}

\begin{figure}[!htbp]
\centering
\includegraphics[width=14cm]{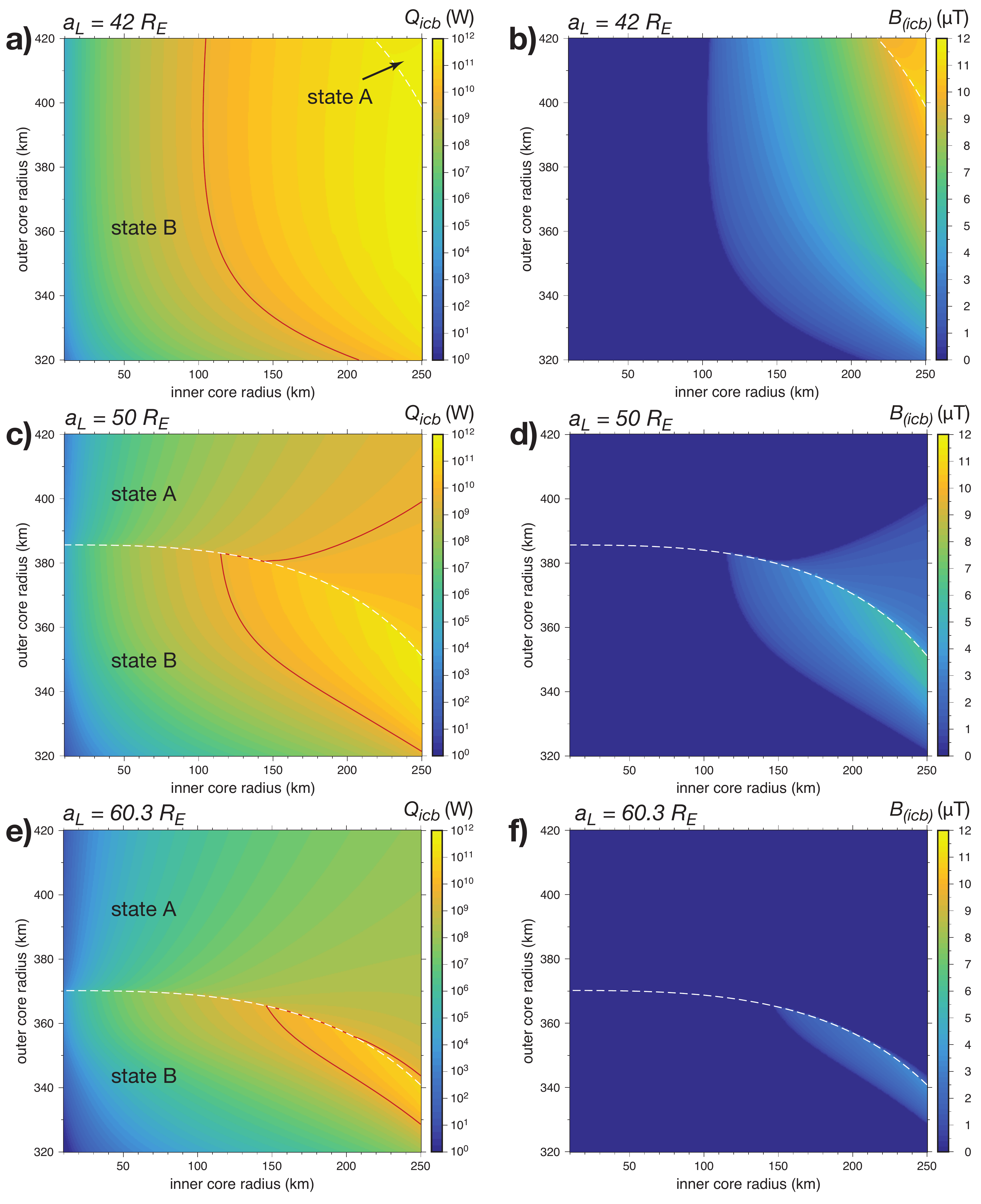}
\caption{\label{fig:QB} Dissipation at the ICB ($Q_{icb}$, in Watts, left column) and magnetic field strength at the lunar surface (${\cal B}_{(icb)}$, in $\mu$T, right column) as a function of $r_s$ and $r_f$, at three different lunar orbit radii: $a_L  = 42 R_E$ (top row), $a_L  = 50 R_E$ (middle row) and  $a_L  = 60.3 R_E$ (bottom row).   $Q_{th} = 4.7 \times 10^9$ W is shown by the red line on the $Q_{icb}$ plots.  The white dashed lines show where the FICN frequency is equal to the Cassini transition frequency.}
\end{figure}

Figure \ref{fig:QB} further illustrates how the interior density structure of the Moon affects $Q_{icb}$ and ${\cal B}_{(icb)}$.  It shows how $Q_{icb}$ and ${\cal B}_{(icb)}$ vary as a function inner core radius $r_s$ and fluid core radius $r_f$, at three specific choices of lunar orbit radius, or equivalently three specific epochs. These are snapshots in time so are independent of the history of inner core growth.  The three epochs that are chosen are: $a_L=42 R_E$ (panels a-b), approximately where $Q_{icb}$ becomes larger than $Q_{cmb}$ in Figure \ref{fig:Qicb2}, and coinciding also to when $Q_{icb}$ reaches its largest value; $a_L=50 R_E$ (panels c-d), coinciding approximately with the shutoff of the dynamo from differential precession at the CMB; and $a_L=60.3 R_E$ (panels e-f), corresponding to today.  
 
As expected, $Q_{icb}$ generally increases with inner core size.  However, the complete picture is more intricate, as $Q_{icb}$ also depends on the Cassini state occupied by the inner core and how close the FICN frequency $\Omega_{ficn}$ is to Cassini transition frequency $\Omega_t$.  The discontinuity in the $Q_{icb}$ contours (identified by a white dashed line) marks the location in the $r_s-r_f$ space where $\Omega_{ficn}$ is equal to $\Omega_t$ ($\Omega_t$ is equal to $2\pi/19.6$ yr$^{-1}$, $2\pi/19.3$ yr$^{-1}$ and $2\pi/16.4$ yr$^{-1}$ for $a_L = 42 R_E$, $50 R_E$ and $60.3 R_E$, respectively).  This discontinuity marks the boundary between models for which the inner core is in Cassini state A ($r_f$ values above the discontinuity) versus those in state B ($r_f$ values below the discontinuity).  For a given combination of $r_s$ and $r_f$, the closer $\Omega_{ficn}$ is to $\Omega_t$, the larger $Q_{icb}$ is.  The largest absolute inner core tilt angles, and thus the largest $Q_{icb}$, are achieved in state B.  $Q_{icb}$ amplitudes are highest at $a_L=42 R_E$ and decrease with increasing $a_L$,  consistent with the behaviour shown in Figure  \ref{fig:Qicb2}.
  
The red contour line on the $Q_{icb}$ panels corresponds to $Q_{th}=4.7 \times 10^9$ W, our chosen threshold for dynamo action.  For $Q_{icb}> Q_{th}$, a magnetic field is generated, and its strength at the lunar surface, ${\cal B}_{(icb)}$, is shown in panels b-d-f of Figure \ref{fig:QB}.  The largest magnetic field strengths coincide with the largest values of $Q_{icb}$.  Being mindful of the caveats on our estimates of the magnetic field strength, the maximum ${\cal B}_{(icb)}$ at $a_L=42 R_E$, $50 R_E$ and $60.3R_E$ are, respectively, $11.4 \mu$T, $6.4 \mu$T and $3.4 \mu$T.

One conclusion that emerges from Figure \ref{fig:QB} is that the inner core of the Moon must be sufficiently large to sustain a dynamo by differential precession at the ICB. At $a_L=42 R_E$, the minimum inner core size is approximately 100 km, and the requirement on inner core size increases with $a_L$.  A second conclusion from Figure \ref{fig:QB} is that at a given epoch $Q_{icb}$ is above $Q_{th}$ only for a specific range of core density models, and this range gets narrower the further away the Moon is from Earth. 

Interestingly, Figure \ref{fig:QB}f suggests that there is still a range of $r_s$ and $r_f$ that allow for dynamo action today, more specifically models for which the inner core is larger than approximately 150 km, is in Cassini state B and with a FICN frequency close to the Cassini transition frequency.  Obviously, this depends directly on our assumption of a power threshold of $Q_{th}=4.7 \times 10^9$ W.  A higher threshold would further restrict the range of models for which a dynamo is possible; a weaker threshold would extend it.  But if the threshold that we have used is approximately correct, the fact that the Moon does not have an on-going dynamo today implies then that: 1) the inner core radius is smaller than approximately 150 km; or 2) that the inner core currently occupies state A; or 3) the inner core is in state B, but the FICN frequency $\Omega_{ficn}$ is not close to the present-day transition frequency  of $\Omega_t = 2\pi/16.4$ yr$^{-1}$ ($\Omega_{ficn} > 2\pi/19$ yr$^{-1}$  for $r_s=200$ km; $\Omega_{ficn} > 2\pi/20$ yr$^{-1}$  for $r_s=250$ km).  If the present-day Moon has an inner core radius larger than 150 km, and if its FICN frequency is close but outside the interval $2\pi/16.4 - 2\pi/20$ yr$^{-1}$, a lunar dynamo powered by precession at the ICB may have shut down only very recently. 

\begin{figure}[!htbp]
\centering
\includegraphics[width=7cm]{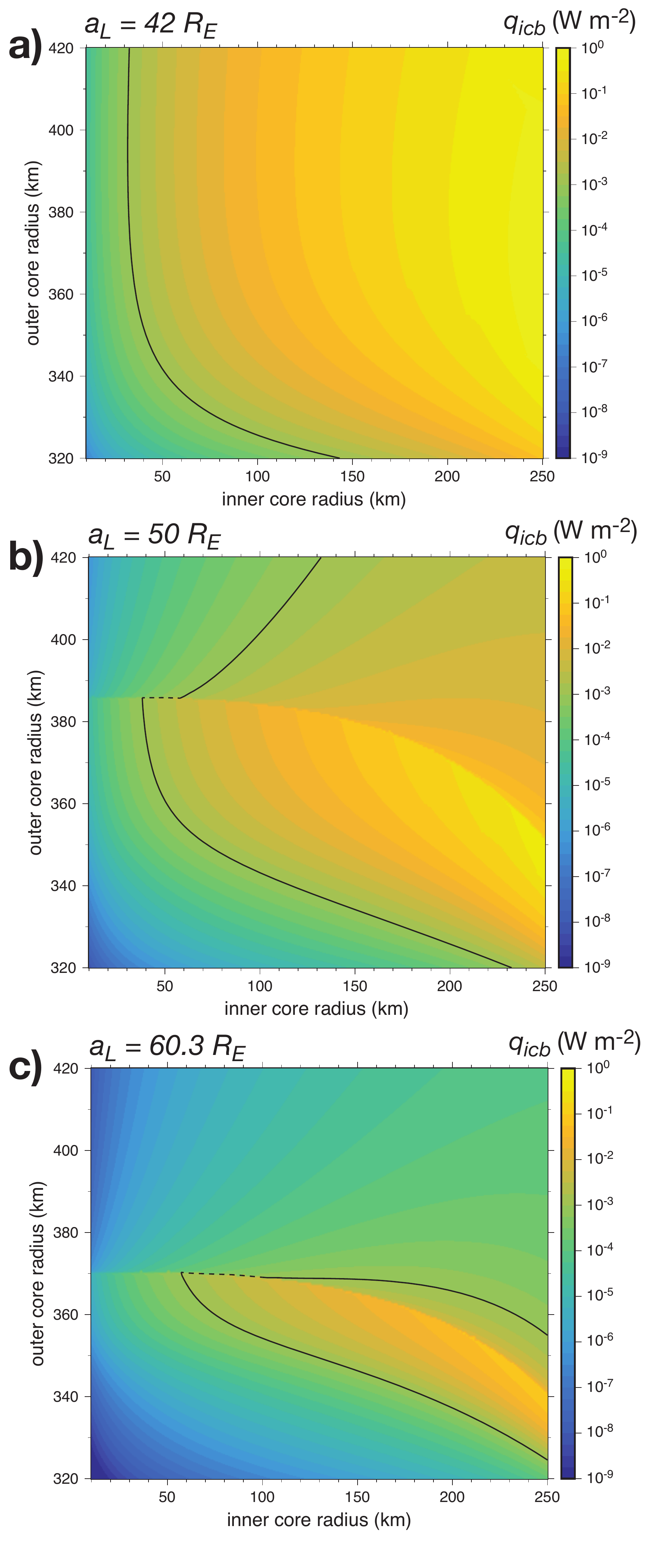}
\caption{\label{fig:QF} Heat flux from viscous dissipation at the ICB from ($q_{icb}$, in W m$^{-2}$) as a function of $r_s$ and $r_f$ for a) $a_L  = 42 R_E$, b) $a_L  = 50 R_E$, and  c) $a_L  = 60.3 R_E$.  Black lines identify contours for which $q_{icb}$ = 1 mW m$^{-2}$.}
\end{figure}

The viscous dissipation at the ICB shown in the $Q_{icb}$ panels on Figure \ref{fig:QB} constitutes a source of heat at the bottom of the fluid core.  It is this heat that can drive thermal convection and generate dynamo action.  Figure \ref{fig:QF} shows the associated heat flux at the ICB, $q_{icb}$, as a function of $r_s$ and $r_f$, calculated from 

\begin{equation}
q_{icb} = \frac{Q_{icb}}{4\pi r_s^2} \, .
\end{equation}
At $a_L = 42R_E$ and $50R_E$, a large portion of the $r_s-r_f$ model space features a heat flux larger than 1 mW/m$^2$.  The heat flux is considerably weaker today ($a_L = 60.3R_E$) although there is still a large portion of the $r_s-r_f$ space where it is above 0.1 mW/m$^2$.  For comparison, typical adiabatic heat flux values for the lunar core are of the order of 3-10 mW/m$^2$ \cite[][]{dwyer11,laneuville14} and the heat flux associated with the latent heat release from inner core crystallization is of the order of 2 mW/m$^2$ \cite[e.g.][]{laneuville14}.  Hence, in addition to considerations of the past dynamo of the Moon, the amplitude of the viscous dissipation at the ICB from the differential precession between the fluid and solid cores provides an important source of heat which cannot be neglected in thermal evolution models.  

Lastly, the $Q_{icb}$, ${\cal B}_{(icb)}$ and  $q_{icb}$ contour maps of Figures  \ref{fig:QB} and \ref{fig:QF} are tied to the choices we have made for the density ($\rho_c$) and thickness ($h$) of the crust and for the density of the inner core ($\rho_s$).  These choices influence the densities of the mantle and fluid core that can match $I_{sm}$ and $\bar{\rho}$, and in turn, this affects the frequency of the FICN for a given combination of $r_s$ and $r_f$.  With different assumptions on $h$, $\rho_c$ and $\rho_s$, the contours of $Q_{icb}$, ${\cal B}_{(icb)}$ and  $q_{icb}$ would be shifted in $r_s-r_f$ space.  Our general conclusions remain unaltered, but one should be careful in extracting specific values of $Q_{icb}$, ${\cal B}_{(icb)}$ and  $q_{icb}$ as a function of $r_s$ and $r_f$ from Figures \ref{fig:QB} and \ref{fig:QF}.

\section{Discussion and Conclusions}

In agreement with \cite{dwyer11}, we find that the dissipation at the CMB of the Moon, $Q_{cmb}$, from viscous friction generated by the differential precession between the mantle and fluid core was large in the past, reaching values above $10^{13}$ W when the Moon was orbiting closer to Earth.  Our estimates of $Q_{cmb}$ differ slightly from those presented in \cite{dwyer11} because we have used an updated measure of the present-day turbulent dissipation at the CMB, and also because our rotational model takes into account the misalignment of the rotation vector of the fluid core with respect to the ecliptic normal.  Using the power required to keep the fluid core in an adiabatic state as a guideline for the threshold of dynamo action, $Q_{cmb}$ was sufficient to power a dynamo when the Moon was closer to Earth than approximately $50 R_E$.  Given the lack of scaling laws associated with precession driven dynamos, obtaining estimates of the magnetic field strength from such a dynamo is difficult.  Based on a scaling law derived from convective dynamos, the strength of the lunar surface magnetic field may have been as high as $20-30 \mu$T, although this is likely an overestimate as we explain further ahead.  The amplitude of $Q_{cmb}$, the magnetic field strength associated with it, and the precise timing of the dynamo shutoff are only weakly dependent on the density structure of the lunar core and the presence of an inner core.

The novel contribution from our study is the computation of estimates of the dissipation from turbulent friction at the ICB of the Moon, $Q_{icb}$, caused by the differential precession between the inner core and fluid core.  We find that $Q_{icb}$ can be as high as approximately $10^{10}$ W in the present-day Moon for an inner core radius of approximately 200 km and for a tight window of fluid core density models.  $Q_{icb}$ may have reached amplitudes in the range of $10^{10}-10^{11}$ W in the past for a broad range of core density models.  The largest source of error in these calculations is the unknown numerical constant involved in our viscous coupling model, and our estimates of $Q_{icb}$ may be errant by an order of magnitude.

The amplitude of $Q_{icb}$ and its evolution through time are highly sensitive to the interior density structure of the core, notably the size of the inner core, but also the density contrast at the ICB.  Both are influencing the FICN frequency, and it is how the latter compares with the precession frequency that determines the tilt angle of the inner core in its Cassini state (SD18).  For sensible density models of the core, the FICN frequency has remained within the resonant band of the precession frequency for the whole range of orbital radius that we have covered in our study ($>34R_E$) and large inner core tilt angles with respect to the mantle result by resonant excitation.  The closer the FICN frequency is to the transition frequency between Cassini states A and B of the inner core, the larger the inner core tilt angle is.  Inner core tilt angles, and consequently $Q_{icb}$, are typically larger when the inner core occupies state B.  

The heat generated by the differential precession between the solid and fluid core is released at the ICB. This heat may drive a dynamo by thermal convection in the fluid core provided $Q_{icb}$ is larger than the adiabatic heat flow out of the core.  This power threshold is likely a lower bound because flows mechanically forced by the precession of the inner core can potentially further assist dynamo action.  On the basis of this threshold, we have shown that $Q_{icb}$ may have been sufficiently large in the past to power a lunar dynamo.  A key requirement is that the inner core radius must be larger than approximately 100 km.   Interestingly, $Q_{icb}$ can remain above the dynamo threshold for a lunar orbit radius larger than $50 R_E$, and thus can outlive a dynamo generated by differential precession at the CMB.  In fact, a range of core density models are compatible with a dynamo persisting until very recently.   The magnetic field amplitude at the lunar surface that we predict from such a dynamo is of the order of a few $\mu$T.  Although this should be viewed as an order of magnitude estimate at best, it is nevertheless compatible with the lunar paleointensities recorded after 3 Ga \cite[e.g.][]{weiss14}.  

We have presented our results in terms of lunar orbit radius $a_L$ and have not attempted to connect $a_L$ to a specific time-history.  A series of models relating $a_L$ to time before present \cite[][]{ooe90,walker83,webb82} are summarized in \cite{dwyer11}.  Differences between them are important, but we can use the mean of these models \cite[presented in Figure S2 of][]{dwyer11} to obtain approximate yardsticks.  The smaller $a_L$ that we considered, $34 R_E$, corresponds to approximately 4.2 Ga; $a_L=42 R_E$, the point beyond which $Q_{icb}$ can exceed $Q_{cmb}$, corresponds to approximately 3.5 Ga; and $a_L=50 R_E$, the shutoff point of the dynamo from precession at the CMB, corresponds to approximately 2.2 Ga. Using this rough mapping, the large ($>20 \mu$T) field strengths produced by precession between the mantle and the fluid core are consistent (although weaker) with the paleointensities of the high-field epoch between 4.2-3.5 Ga \cite[e.g.][]{weiss14}.  This mantle driven precession dynamo would operate until 2.2 Ga, though with weaker field intensities, consistent with the paleointensities of the  weak-field epoch.  This was the main conclusion of \cite{dwyer11}.

Complementing this picture, our results show that a dynamo driven by the precession of the inner core can outlive the dynamo driven by mantle precession.  Hence, not only this inner core driven precession dynamo may further explain a part of the lunar magnetic field recorded in the weak-field epoch, it can also explain a lunar dynamo persisting until as recently as 1 Ga \cite[e.g.][]{mighani20}, long after the mantle driven precession dynamo would have shutoff.  Hence, a combination of precession driven dynamos, by the mantle and the inner core, may be consistent with at least a part of the lunar paleomagnetic record.  Obviously, this does not preclude that a dynamo driven by thermo-chemical convection in the core \cite[e.g.][]{laneuville14} or the lower mantle \cite[e.g.][]{scheinberg18} may have coexisted with these precessionally driven dynamos.

The onset time of the dynamo driven by inner core precession depends on the history of inner core growth, so it may have been delayed by a fraction to a couple of Gyr after the Moon formed. This dynamo ceased when $Q_{icb}$ eventually dropped below the power threshold to maintain the core adiabat.  This may have occurred smoothly as the lunar orbit evolved, or it may be connected to a transition from state B to state A of the Cassini state occupied by the inner core.  The tilt angle of the inner core is smaller in state A, and hence $Q_{icb}$ would have dropped significantly after such a transition, though large scale flows and a possibly more energetic temporary dynamo may have resulted in the process of the transition.   Perhaps offering support for such a scenario, the most recent estimate of the CMB radius from LLR is $381\pm12$ km \cite[][]{viswanathan19}: this places the inner core in Cassini state A at present-day (see Figure \ref{fig:QB}ef) but relatively close to state B and consistent with a recent transition.

There are important feedback effects that we have not taken into account in our rotational model which may significantly alter our results. First, once a magnetic field is present in the lunar core, electromagnetic (EM) coupling at the ICB acts to reduce the differential rotation between the fluid and solid cores (see for instance DW16).  On the one hand, the reduced differential rotation at the ICB implies a weaker viscous dissipation.  On the other hand, the shearing of the radial magnetic field at the ICB would introduce EM dissipation.  Ultimately, the source of energy remains the amplitude of the differential precession at the ICB in the absence of a dynamo.   Hence, the total of the viscous and EM dissipation may not be wholly different from the $Q_{icb}$ values we have estimated, just separated in different pools.  Without an actual theoretical or numerical model it is difficult to predict precisely how taking into account EM coupling would alter the estimates of $Q_{icb}$ that we have presented.  

A second feedback that we have neglected is how viscous friction at the CMB and ICB may alter the mutual orientations of the rotation vectors of the mantle, fluid core and inner core.  The Cassini state model of SD18 that we have used assumes no dissipation.  However, the large viscous friction at the CMB and ICB, especially at earlier times in the lunar history, may limit the misalignment between the different rotation vectors.  

A third feedback effect that is missing is how $Q_{icb}$ and $Q_{cmb}$ may alter the evolution of $I$ and $\Omega_p$ as a function of orbit radius.  Viscous friction at the CMB and ICB of the Moon act to dissipate the rotational and orbital energy of the Moon.  As an example, a rate of energy dissipation of $Q$ within the Moon, regardless of its nature, leads to a reduction of the inclination $I$ of the lunar orbit according to \cite[e.g. Equation (17) of][]{chen16}

\begin{equation}
\frac{d I}{dt} = - \frac{1}{\tan I} \frac{a_L \, Q}{GM_E M}  \, ,
\end{equation}
where $M$ and $M_E$ are the masses of the Moon and Earth and $G$ is the gravitational constant.  For small $I$, a typical attenuation timescale $\tau_I$ of $I$ is then

\begin{equation}
\tau_I \approx  \frac{I^2 GM_E M}{a_L \,Q} \, .
\end{equation}
To give an estimate of $\tau_I$, let us use as a guide an inclination of $I\approx10^\circ$ at $a_L=40 R_E$.  For $Q=10^{12}$ W, this gives $\tau_I \approx $ 100 Myr.  Energy dissipation of this magnitude inside the Moon should have rapidly driven the inclination close to zero and this is inconsistent with a present-day residual value of $I=5.145^\circ$.  This simple order of magnitude estimate neglects the rotation of the Earth in the energy and angular momentum budgets and also neglects the possibility of inclination re-excitation events \cite[e.g][]{pahlevan15}.  Nevertheless, it illustrates that some of the high dissipation values that we predict are likely not compatible with the time-history of $I$ that we have used.  In order to lengthen the attenuation timescale $\tau_I$ to a more realistic estimate of 1 Gyr, viscous dissipation in the core of the Moon should be limited to approximately $10^{11}$ W.  
 
This problem was pointed by \cite{dwyer11} as they realized that the very large $Q_{cmb}$ exceeding  $3 \times 10^{11}$ W (when $a_L<43 R_E$ in our Figure 2) would lead to widespread mantle melting, suggesting that such large dissipation never occurred.  If so, then the large magnetic field amplitudes at the surface in excess of 10 $\mu$T at earlier times are likely also overestimated.  This makes it more difficult to explain the lunar paleointensities of the order of 100 $\mu$T in the high-field epoch by a precession dynamo.  Likewise, viscous dissipation at the ICB in excess of $10^{11}$ W that we have calculated in our study are unlikely realistic.  As illustrated by Figure \ref{fig:QB}, predictions of $Q_{icb}$ larger than  $10^{11}$ W are associated with an inner core radius larger than 200 km.  Hence, this limits the validity of our results to inner core radii smaller than 200 km.  Our general conclusions are not altered, except that the largest surface magnetic field amplitudes that we predict are limited to approximately 5 $\mu$T.

Ideally, viscous dissipation at the CMB and ICB should be included in calculations of the orbital evolution of the Moon.  The power available to drive a dynamo, either mechanically of thermally, could thus be estimated in a self-consistent manner.  Recent efforts have been made in this direction \cite[e.g.][]{cuk19} although in a limited way and in the absence of an inner core, as challenges remain important. If viscous dissipation at the ICB throughout the Moon's history must be limited so that it is is not in conflict with the present-day lunar orbit inclination, this could place constraints on the maximum inner core size.   We hope that our study may serve as an additional motivation to include the presence of an inner core and viscous friction at the fluid core boundaries in lunar orbital evolution models.  

Our results indicate that the heat flux $q_{icb}$ associated with $Q_{icb}$ can be of the order of a few mW m$^{-2}$.  This heat flux may drive convective flows and power a dynamo.  But the mantle would still act as a bottleneck for this extra heat.  A higher core temperature decreases the radius at which the adiabatic temperature intersects the melting temperature of the iron alloy, hence the onset and rate of inner core growth is ultimately controlled by how much heat can escape the core.   Thermal evolution models for the core of the Earth and planets typically do not include a contribution drawn from rotational or orbital energy \cite[e.g.][]{nimmo15}.  Tidal heating has been considered in the thermal budget for many moons of the solar system \cite[e.g][]{breuer15,nimmo16}, including the Moon \cite[e.g.][]{peale78,meyer10}, but viscous heating associated with precession is usually ignored.  Our results suggest that the latter may play a first order role in the thermal evolution of the lunar core.  To put our results  into perspective, the latent heat released from inner core crystallization -- the largest contribution to the heat budget in the absence of precession -- is of the order of 2 mW m$^{-2}$ \cite[e.g.][]{laneuville14}.  Once the inner core radius reaches about 50 km, the heat flow from viscous friction connected to the precession of the inner core is of the same order and cannot be neglected.  The rate of inner core growth may then be significantly slowed down by the additional heat at the ICB induced by inner core precession.  A slower inner core growth would further reduce the latent heat released at the ICB and thus further decrease its importance compared to $q_{icb}$.

Lastly,  if the heat released at the ICB is subcritical for a thermally driven dynamo, core flows mechanically forced by the precession of an elliptically shaped inner core may be capable of generating dynamo action by themselves.  Whether this is possible is unknown at present.  Ultimately, the conditions for the onset of such a dynamo, and the form and strength that its associated magnetic field may take inside the core and at the lunar surface, can only be answered by an actual dynamical model.   These questions are particularly relevant given the recent paleomagnetic intensities weaker than $0.1 \mu$T recorded in two Apollo samples and dated at $0.44 \pm 0.01$ and $0.91 \pm 0.11$ Ga \cite[][]{mighani20}.  Since such magnetic field intensities are much smaller than those typically expected from convective dynamos, \cite{mighani20} conclude that the lunar dynamo must have likely ceased before $\sim0.8$ Ga.  However, a magnetic field generated by mechanically forced flows from precession at the ICB could have a r.m.s. strength of the order of $10-100 \mu$T inside the core but with its energy dominantly in small length scales features.  Because of the sharp spatial attenuation of this field outside the core, the larger length scale part would dominate the field recorded at the lunar surface but it may only amount to a fraction of a $\mu$T.   Indeed, numerical models of dynamos generated by precession at the CMB suggest this is the case \cite[][]{cebron19}.  Hence, a lunar dynamo from precession at the ICB generating surface field strengths of a fraction of a $\mu$T may be compatible with the weak paleointensities recorded after 1 Ga, which would push the end of the lunar dynamo to more recently than 0.44 Ga.  We hope that our results may serve as a motivation to modellers to attempt to address these questions.

\acknowledgments
We thank Francis Nimmo and an anonymous reviewer for their constructive comments which helped to improve this paper.  The model used in this research is presented in detail in \cite{stys18}.  Input parameters that are different than those used in Table 1 of  \cite{stys18} are described in the text.  Figures were created using the GMT software \cite[]{gmt}. The source codes, GMT scripts and data files to reproduce all figures are freely accessible at {\em https://doi.org/10.7939/DVN/T1DTCM}.  This work was supported by an NSERC/CRSNG Discovery Grant.  


\end{document}